\documentstyle[11pt]{article}
\def\bkR{{\rm I\kern-.17em R}}
\def\bkC{{\rm \kern.24em \vrule width.05em height1.4ex depth-.05ex \kern-.26em C}}

\begin{document}

\title{{\bf Deformation quantization of noncommutative quantum mechanics and dissipation}\footnote{Based on a talk presented by Jo\~ao Prata at D.I.C.E. 2006, Piombino, Italy.} }

\author{{\bf C Bastos$^{1,2}$, O Bertolami$^{1,2}$, N C Dias$^{3,4}$ and J N Prata$^{3,4}$} \\ \\ 
{$^1$ Departamento de F\'{\i}sica, Instituto Superior T\'ecnico, Avenida Rovisco Pais 1,}\\ 
{1049-001 Lisboa, Portugal} \\ 
{$^2$ Centro de F\'{\i}sica dos Plasmas, Instituto Superior T\'ecnico, Avenida Rovisco Pais 1,}\\
{1049-001 Lisboa, Portugal} \\ 
{$^3$ Departamento de Matem\'atica, Universidade Lus\'ofona de Humanidades e Tecnologias,}\\
{Avenida Campo Grande 376, 1749-024 Lisboa, Portugal}\\ 
{$^4$ Grupo de F\'{\i}sica Matem\'atica, Universidade de Lisboa,
Avenida Prof. Gama Pinto 2,}\\{ 1649-003 Lisboa, Portugal}\\ \\
{E-mail: cbastos@fisica.ist.utl.pt, orfeu@cosmos.ist.utl.pt,}\\ {ncdias@mail.telepac.pt, joao.prata@mail.telepac.pt}}

\maketitle

\maketitle

\begin{abstract}
We review the main features of the Weyl-Wigner formulation of noncommutative quantum mechanics. In particular, we present a $\star$-product and a Moyal bracket suitable for this theory as well as the concept of noncommutative Wigner function. The properties of these quasi-distributions are discussed as well as their relation to the sets of ordinary Wigner functions and positive Liouville probability densities. Based on these notions we propose criteria for assessing whether a commutative regime has emerged in the realm of noncommutative quantum mechanics. To induce this noncommutative-commutative transition, we couple a particle to an external bath of oscillators. The master equation for the Brownian particle is deduced.
\end{abstract}

\section{Introduction}

Recently, there has been a great interest in field theories on noncommutative space-times \cite{Seiberg, Douglas}. Usually, in these approaches one assumes that time remains a commutative parameter to avert problems with the lack of unitarity or causality. The nonrelativistic one-particle sector of such theories with canonical commutation relations is denoted by noncommutative quantum mechanics (NCQM) (see \cite{Orfeu} and references therein). This theory is characterized by a so-called {\it extended Heisenberg algebra}. For a $d$-dimensional space the position and momentum variables are collectively written as $z=(q,p)$. In the sequel Latin letters run from $1$ to $d$ (e.g. $i,j,k =1, \cdots, d$), whereas Greek letters stand for phase space indices (e.g. $\alpha, \beta, \gamma =1, \cdots, 2d$). Moreover, summation over repeated indices is understood. The extended Heisenberg algebra then reads:  
\begin{equation}
\left[\hat z_{\alpha}, \hat z_{\beta} \right]= i \hbar \Omega_{\alpha \beta}, \hspace{1 cm} \alpha, \beta = 1, \cdots, 2d,
\label{Eq1.1}
\end{equation}
where ${\bf \Omega}$ is a real constant antisymmetric matrix with $\det {\bf \Omega} >0$, which can be cast in the block form:
\begin{equation}
{\bf \Omega} = \left(
\begin{array}{c c}
{\bf \Theta} & {\bf I}_{d \times d}\\
- {\bf I}_{d \times d} & \overline{{\bf \Theta}} 
\end{array}
\right).
\label{Eq1.2}
\end{equation}
Here ${\bf I}_{d \times d}$ denotes the $(d \times d)$ identity matrix and ${\bf \Theta}, \overline{{\bf \Theta}}$ are real antisymmetric matrices with entries $\theta_{ij}, \overline{\theta}_{ij}$, $(i,j=1, \cdots, d)$, measuring the noncommutativity, which is assumed to be small, in the spatial and momentum sectors, respectively. The extended Heisenberg algebra is isomorphic with the ordinary Heisenberg algebra $\xi = (R, \Pi)$
\begin{equation}
\left[\hat{\xi}_{\alpha}, \hat{\xi}_{\beta} \right] = i \hbar J_{\alpha \beta} , \hspace{1 cm}
{\bf J} = \left(
\begin{array}{c c}
{\bf O}_{d \times d} & {\bf I}_{d \times d} \\
- {\bf I}_{d \times d} & {\bf O}_{d \times d}
\end{array}
\right)
\label{Eq1.3}
\end{equation}
via the so-called Seiberg-Witten (SW) map, \cite{Seiberg}, a linear transformation of the form\footnote{${\bf S}$ is a real constant nonsingular matrix.}
\begin{equation}
z_{\alpha} = S_{\alpha \beta} \xi_{\beta}, \hspace{1 cm} {\bf \Omega} = {\bf S} {\bf J} {\bf S}^T.
\label{Eq1.4}
\end{equation}
Note that this transformation is not unique. The Jacobian of the transformation (\ref{Eq1.4}) 
is $\det {\bf S} = \pm \sqrt{\det {\bf \Omega}}$. In view of this isomorphism, the extended Heisenberg algebra admits a representation in terms of the Hilbert space of ordinary quantum mechanics, i.e. ${\cal H} = L_2 (\bkR^d, dR )$.

\vspace{0.3 cm}
\noindent
The main idea of our work is to study the emergence of ordinary quantum mechanics within the realm of noncommutative quantum mechanics. The strategy to induce this transition is to place the noncommutative system in interaction with an external environment. For reasons which will become apparent in what follows it is more adequate to study this environment-induced transition by resorting to the Weyl-Wigner formulation \cite{Zachos}. This will require suitable adaptations to incorporate the noncommutativity of the position and momentum variables. This work was initiated in \cite{Hu1}. Some proposals for the noncommutative Wigner functions appeared recently in the literature \cite{Jing, Rosenbaum1}. In the present work we wish to present an overview of our work \cite{Bastos1}, \cite{Dias1}, \cite{Dias3} as well as some new results. 

\section{Deformation quantization for noncommutative systems}

Elements $\hat A$ of the universal envelopping algebra $\hat{\cal A} ({\cal H})$ of the Heisenberg variables (\ref{Eq1.3}) or of the vector space $\hat{\cal F}$ of Hilbert-Schmidt operators acting on the Hilbert space ${\cal H}$, can be mapped bijectively to suitable subspaces of the space ${\cal S}' (\bkR^{2d})$ of generalized functions on the phase space $T^*M \simeq \bkR^{2d}$ via the Weyl-Wigner map \cite{Bastos1}:
\begin{equation}
W_{\xi}: \hat A (\hat{\xi}) \longmapsto A_W (\xi) \equiv W_{\xi} (\hat A) = \frac{1}{(2 \pi \hbar)^d} \int d \eta Tr \left[ \hat A e^{i \eta \cdot (\xi - \hat{\xi})} \right].
\label{Eq2.1}
\end{equation}
The phase space can thus be endowed with a noncommutative product and a bracket according to:
\begin{equation}
W_{\xi} (\hat A \cdot \hat B)  \equiv  W_{\xi} (\hat A) \star_{\xi} W_{\xi} (\hat B) , \hspace{0.5 cm} W_{\xi} (\left[\hat A, \hat B \right] )  \equiv  i \hbar \left[ W_{\xi} (\hat A) , W_{\xi} (\hat B ) \right]_{\xi},
\label{Eq2.2}
\end{equation}
commonly known as the Moyal $\star$-product and the Moyal bracket. From Eqs. (\ref{Eq2.1}) and (\ref{Eq2.2}) one can write the Moyal product and the Moyal bracket as:
\begin{eqnarray}
A(\xi) \star_{\xi} B (\xi) & = & A(\xi) \exp \left( \frac{i\hbar}{2} \buildrel{\leftarrow}\over\partial_{\xi_{\alpha}} J_{\alpha \beta} \buildrel{\rightarrow}\over\partial_{\xi_{\beta}}  \right) B (\xi)~,
\label{Eq2.3}\\
\left[A(\xi) , B (\xi) \right]_{\xi} & = & \frac{2}{\hbar} A(\xi) \sin \left( \frac{\hbar}{2} \buildrel{\leftarrow}\over\partial_{\xi_{\alpha}} J_{\alpha \beta} \buildrel{\rightarrow}\over\partial_{\xi_{\beta}}  \right) B (\xi)~,
\label{Eq2.4}
\end{eqnarray}
Likewise, the density matrix $\hat{\rho} \in \hat{\cal F}$ is mapped to the so-called Wigner function:
\begin{equation}
f^W (\xi) \equiv \frac{1}{(2 \pi \hbar)^d} W_{\xi} (\hat{\rho}) (\xi).
\label{Eq2.5}
\end{equation}
In particular, for a pure state $\hat{\rho} = | \psi\rangle\langle \psi|$, we may write it as:
\begin{equation}
f^W (R, \Pi) = \frac{1}{(\pi \hbar)^d} \int dy ~e^{-2 i \Pi \cdot y / \hbar} \psi (R + y) \psi^* (R-y).
\label{Eq2.6}
\end{equation}
The question then is how one can implement a Weyl-Wigner formulation for NCQM. The answer lies in the SW map and in a covariant generalization of the Weyl-Wigner map defined in Ref. \cite{Dias2}. This map $W_z^{\xi}$ takes an operator $\hat A (\hat z)$ expressed in terms of the noncommutative variables $\hat z$ to a phase space counterpart $A(z)$ so that the following diagram is commutative\footnote{The transformations $S$ and $\hat S$ are basically the same. They are both given by (\ref{Eq1.4}). However $\hat S$ acts on operators whereas $S$ acts on phase space functions.}:
$$
\begin{array}{l c c}
\hat{A}(\hat z) & ------------- \longrightarrow & \hat{A}'(\hat{\xi}) \equiv \hat A(\hat z(\hat{\xi}))\\
& \hat S &\\
W_z^{\xi} \downarrow & & \downarrow W_{\xi}\\
& & \\
A(z)&------------- \longrightarrow & A' (\xi)=A(z(\xi))\\
& S &\\
\end{array}
$$
where $S$ is the SW transformation (\ref{Eq1.4}). Hence $W_z^{\xi} = S^{-1} \circ W_{\xi} \circ \hat S$. The $\star$-product then changes according to: 
\begin{equation}
A(z) \star_z B (z)  = A(z) \exp \left( \frac{i\hbar}{2} \buildrel{\leftarrow}\over\nabla_{z_{\alpha}} J_{\alpha \beta}' \buildrel{\rightarrow}\over\nabla_{z_{\beta}}  \right) B (z),
\label{Eq2.7}
\end{equation}
 where the covariant derivatives are associated with the Christoffel symbols: $\Gamma_{\alpha \beta \gamma} = {\partial z_{\gamma}\over\partial \xi_{\sigma}}{\partial^2 \xi_{\sigma}\over\partial z_{\alpha} \partial z_{\beta}}$. Since the SW transformation (\ref{Eq1.4}) is linear, the latter vanish and the covariant derivatives reduce to ordinary ones. The transformed symplectic matrix is given by (cf. Eq. (\ref{Eq1.4})): $J_{\alpha \beta}' =  {\partial z_{\alpha}\over\partial \xi_{\gamma}}{\partial z_{\beta}\over\partial \xi_{\sigma}} J_{\gamma \sigma} =  S_{\alpha \gamma} S_{\beta \sigma} J_{\gamma \sigma} = \Omega_{\alpha \beta}$. Altogether, we obtain the extended $\star$-product and the extended Moyal bracket:
\begin{eqnarray}
A(z) \star_z B (z) & = & A(z) \exp \left( \frac{i\hbar}{2} \buildrel{\leftarrow}\over\partial_{z_{\alpha}} \Omega_{\alpha \beta} \buildrel{\rightarrow}\over\partial_{z_{\beta}}  \right) B (z)~,
\label{Eq2.9}\\
\left[A(z) , B (z) \right]_z & = & \frac{2}{\hbar} A(z) \sin \left( \frac{\hbar}{2} \buildrel{\leftarrow}\over\partial_{z_{\alpha}} \Omega_{\alpha \beta} \buildrel{\rightarrow}\over\partial_{z_{\beta}}  \right) B (z)~.
\label{Eq2.10}
\end{eqnarray}
Likewise, from the extended Weyl-Wigner transform $W_z^{\xi}$ we can define the counterpart of the Wigner function for NCQM. We shall call it the noncommutative Wigner function (NCWF):
\begin{equation}
f^{NC} (z) \equiv \frac{1}{(2 \pi \hbar)^d \sqrt{\det {\bf \Omega}}} W_z^{\xi} (\hat{\rho}) (z)= \frac{1}{\sqrt{\det {\bf \Omega}}} f^W (\xi (z)),
\label{Eq2.11}
\end{equation}
where obviously $f^W (\xi) = ( 2 \pi \hbar)^{- d} W_{\xi} (\hat{\rho}) (\xi)$. The determinant of ${\bf \Omega}$, which is a constant, is included to ensure the normalization of the NCWF. 

To obtain the dynamics of the NCWF, one applies the extended Weyl-Wigner transform $W_z^{\xi}$ to the von Neumann equation. One obtains the extended von Neumann-Moyal equation:
\begin{equation}
\frac{\partial}{\partial t} f^{NC} (z,t) = \left[H(z), f^{NC} (z,t) \right]_z, \hspace{0.5 cm} H(z) = W_z^{\xi} (\hat H).
\label{Eq2.12}
\end{equation}
In the next section we shall apply this formulation to an ensemble constituted of a Brownian particle and a bath of oscillators on the plane with spatial noncommutativity. This simplifies considerably our algebra: ${\bf \Theta} = \frac{\theta}{\hbar} {\bf E}$, $\overline{{\bf \Theta}} ={\bf O}$, where ${\bf E}$ is the $2 \times 2$ matrix with entries $\epsilon_{11}=\epsilon_{22}=0$, $\epsilon_{12} = - \epsilon_{21} =1$ and $\theta$ is the only parameter measuring the noncommutativity of the theory. A useful SW map for this extended Heisenberg algebra is:
\begin{equation}
{\bf S} = \left(
\begin{array}{c c}
{\bf I}_{2 \times 2} & -\frac{1}{2} {\bf \Theta}\\
{\bf O}_{2 \times 2} & {\bf I}_{2 \times 2}
\end{array}
\right), \hspace{1 cm} \det {\bf S} =1.
\label{Eq2.13}
\end{equation}
Upon substitution of (\ref{Eq2.13}) into (\ref{Eq2.6}) and (\ref{Eq2.11}) one obtains after some algebra the result of Ref. \cite{Jing}:
\begin{equation}
f^{NC} (q,p) = \frac{1}{(\pi \hbar)^2} \int dy ~ e^{-2 i y \cdot p / \hbar} \psi (q+y) \star_{\theta} \psi^* (q-y),
\label{Eq2.14}
\end{equation}
where the $\star_{\theta}$-product only involves the position variables and is given by: 
\begin{equation}
A(q) \star_{\theta} B (q) =  A(q) e^{ \frac{i \theta}{2} \frac{\buildrel{\leftarrow}\over\partial}{\partial_{q_i}} \epsilon_{ij} \frac{\buildrel{\rightarrow}\over\partial}{\partial_{q_j}}} B (q).
\label{Eq2.15}
\end{equation}
Notice the similarity between the expressions (\ref{Eq2.6}) and (\ref{Eq2.14}). The spatial noncommutativity is encoded in the $\star_{\theta}$-product. The expression (\ref{Eq2.14}) corresponds to the NCWF of a pure state. Mixed states are as usual obtained by constructing convex combinations of states of the form (\ref{Eq2.14}). One can then prove the following constraint on the NCWF:
\begin{equation}
\int d z ~ f^2 (z) 
\left\{
\begin{array}{l l}
= \frac{1}{(2 \pi \hbar)^d} & \mbox{for pure states}	\\
< \frac{1}{(2 \pi \hbar)^d} & \mbox{for mixed states}~.	
\end{array}
\right.
\label{Eq2.16}
\end{equation}
This bound is shared by ordinary Wigner functions. However, if one computes the position marginal distribution of (\ref{Eq2.14}), one obtains:
\begin{equation}
{\cal P} (q) \equiv \int dp~f^{NC} (q,p) = \psi (q) \star_{\theta} \psi^* (q).
\label{Eq2.17}
\end{equation}
Unlike the marginal distribution of the Wigner function, this is not a true probability density, as the nonlocal $\star_{\theta}$ spoils, in general, its positivity. Moreover, since ${\cal P} (q)$ can be regarded as a positive, normalized element of the $\star_{\theta}$ algebra, one can get an additional constraint on the NCWF:
\begin{equation}   
\int dq~ {\cal P}^2 (q) = \int dq \int dp \int dp'~ f^{NC} (q,p) f^{NC} (q,p') \le \frac{1}{2 \pi \theta}~,
\label{Eq2.18}
\end{equation}
which has no counterpart in ordinary quantum mechanics. So this constitutes a measure of noncommutativity. Let us denote by ${\cal W}$, ${\cal W}^{NC}$ and ${\cal L}$ the sets of ordinary Wigner functions, NCWF and positive Liouville measures, respectively. Using properties such as (\ref{Eq2.16})-(\ref{Eq2.18}), one can construct explicitly families of distributions in ${\cal W}$, ${\cal W}^{NC}$, ${\cal L}$, and prove that all the following sets are nonempty:
\begin{eqnarray}
\Omega_1 = {\cal W} \backslash ( {\cal W}^{NC} \cup {\cal L} ), & \Omega_2 = {\cal W}^{NC} \backslash ( {\cal W} \cup {\cal L} ) , & \Omega_3 = {\cal L} \backslash ({\cal W} \cup {\cal W}^{NC} ), \nonumber\\
\Omega_4 = ({\cal W} \cap  {\cal W}^{NC}) \backslash {\cal L},  & \Omega_5 = ({\cal W} \cap {\cal L} ) \backslash {\cal W}^{NC} , & \Omega_6 = ({\cal W}^{NC} \cap {\cal L} ) \backslash {\cal W}, \nonumber \\
\Omega_7 =  {\cal W} \cap {\cal W}^{NC} \cap {\cal L}.  \label{Eq2.19}
\end{eqnarray}
This analysis allows one to propose the necessary conditions for the noncommutative-commutative transition to take place (this will be the subject of the next section). In the context of environment-induced decoherence \cite{Zeh}, one usually assumes that the quantum-classical transition occurs, once the reduced density matrix for the Brownian particle becomes diagonal in the position basis. A necessary, albeit not sufficient, condition for this to happen is that the associated Wigner function of the Brownian particle becomes everywhere non-negative. In other words, one may say that the Wigner function of the Brownian particle $f_B^W$ which belongs to the set ${\cal W}$ enters into the set ${\cal L}$ of positive Liouville measures at the time $t_{Q-Cl}$ of the quantum-classical transition \cite{Diosi}, i.e.: $f_B^W (z,t) \in {\cal W} \cap {\cal L}$, for $t \ge t_{Q-Cl}$. In analogy with this, one may thus propose the following condition for assessing whether the noncommutative-commutative transition has taken place. If $f_B^{NC}$ is the NCWF of the Brownian particle and $t_{NC-C}$ is the instant when the transition occurs, then:
\begin{equation}
f_B^{NC} (z,t) \in {\cal W} \cap {\cal W}^{NC} , \hspace{0.5 cm} t \ge t_{NC-C}.
\label{Eq2.20}
\end{equation}
 
\section{The noncommutative Hu-Paz-Zhang equation}

As mentioned in the introduction, the strategy to induce the NC-C transition consists in coupling a noncommutative Brownian oscillator to an external reservoir of noncommutative oscillators at thermal equilibrium and thus treat it as an open system \cite{Caldeira}. We shall assume that the coupling $(C_n)$ to be weak so that only the linear response is considered. The Hamiltonian reads:
\begin{equation}
\hat H = \frac{\hat p^2}{2 M} + \frac{1}{2} M \Omega^2 \hat q^2 + \sum_n \left( \frac{ \left(\hat p^{(n)} \right)^2}{2 m_n} + \frac{1}{2} m_n \omega_n^2 \left(\hat q^{(n)} \right)^2 \right) + \sum_n C_n \hat q \cdot \hat q^{(n)}, 
\label{Eq3.1}
\end{equation}
where $\hat q= (\hat q_1 , \hat q_2)$, $\hat p= (\hat p_1 , \hat p_2)$ are the position and momentum of the Brownian oscillator of mass $M$ and bare frequency $\Omega$, $\hat q^{(n)}= (\hat q_1^{(n)} , \hat q_2^{(n)})$, $\hat p^{(n)}= (\hat p_1^{(n)} , \hat p_2^{(n)})$ are the positions and momenta of the bath oscillators with masses $m_n$ and frequencies $\omega_n$.

\noindent
The dynamics of the noncommutative Wigner function $f^{NC} \left(z, \left\{z^{(n)} \right\} \right)$ for the composite system particle plus environment, is governed by the extended von Neumann-Moyal equation (\ref{Eq2.12}). We are interested in the master equation for the reduced noncommutative Wigner function of the Brownian particle, which is obtained by integrating out the environment's degrees of freedom:
\begin{equation}
W(z)  \equiv \int \left( \Pi_n dz^{(n)} \right) f^{NC} \left(z, \left\{z^{(n)} \right\} \right). \label{Eq3.2}
\end{equation}
To simplify the derivation we assume that the initial distributions of the Brownian particle $(W)$ and of the bath $(W^b)$ are uncorrelated, and that the bath is at thermal equilibrium \cite{Dias1}: 
\begin{equation}
W^b \left(\left\{  z^{(n)}  \right\} , t=0  \right) = \prod_n N_n 
\times \exp \left[- a_n \left(p^{(n)} \right)^2 - c_n  \left(q^{(n)} \right)^2 - 2 b_n  L^{(n)} \right], \label{Eq3.3}
\end{equation}
where $L^{(n)} =  q^{(n)} \cdot {\bf E} p^{(n)} = q_i^{(n)} \epsilon_{ij} p_j^{(n)}$ and $a_n, b_n, c_n, N_n$ are certain temperature dependent positive coefficients. After a lengthy computation which follows closely the method of Halliwell and Yu \cite{Halliwell} we arrive \cite{Dias1} at the noncommutative extension of the Hu-Paz-Zhang master equation \cite{Hu2}: 
\begin{equation}
\begin{array}{c}
{\partial W\over \partial t} = - {p\over M} \cdot \nabla_q W + M \Omega^2 q \cdot \nabla_p W + \left( \nabla_p W \right) \cdot {\bf A}(t) q + \nabla_p  \cdot \left( {\bf B}(t) p W \right) + \\
\\
+ \nabla_p  \cdot \left( {\bf C}(t) \nabla_q W \right) + \nabla_p  \cdot \left( {\bf D}(t) \nabla_p W \right) + \frac{\theta}{\hbar}  M \Omega^2 q  \cdot {\bf E}  \nabla_q W -  {\theta\over \hbar}  \nabla_q \cdot {\bf E} \left({\bf A}(t) q W \right) \\
\\
-  {\theta\over\hbar}  \nabla_q \cdot {\bf E} \left({\bf B}(t) p W \right) -  {\theta\over\hbar}  \nabla_q \cdot {\bf E} \left({\bf C}(t) \nabla_q W \right) -  {\theta\over\hbar}  \nabla_q \cdot {\bf E} \left({\bf D}(t) \nabla_p W_r \right),
\end{array} \label{Eq3.4}
\end{equation}
where ${\bf A}, {\bf B}, {\bf C}, {\bf D}$ are time dependent $(2 \times 2)$ matrices. Their explicit form depends solely on the so-called dissipation and noise kernels. In the Heisenberg picture the equation of motion for the Brownian particle is the noncommutative Langevin equation:
\begin{equation}
\ddot Q_i (t) + \Omega^2 Q_i (t) - \frac{\theta}{\hbar} M \Omega^2 \epsilon_{ij} \dot Q_j (t)  
+  \frac{2}{M} \int_0^t ds \hspace{0.2 cm} \eta_{kj} (t-s) \left(\delta_{ik} - \frac{M \theta}{\hbar} \epsilon_{ik} \frac{d}{ds} \right) Q_j (s) = \frac{f_i (t)}{M}, 
\label{Eq3.5}
\end{equation}
where $Q_i  = q_i + \frac{\theta}{\hbar} \epsilon_{ij} p_j$ is used for convenience. The dissipation matrix kernel $\eta_{ij} (t) = \frac{d}{dt} \gamma_{ij} (t)$ is given by:
\begin{equation}
\gamma_{ij} (t) = \int_0^{+ \infty} \frac{d \omega}{\omega} \left[ \delta_{ij} I^{(+)} (\omega)  \cos (\omega t) + \epsilon_{ij} I^{(-)} (\omega)  \sin (\omega t) \right]. \label{Eq3.6}
\end{equation}
The spectral densities read:
\begin{equation}
I^{( \pm)} (\omega) = \sum_n \frac{C_n^2 \omega^2 }{4 m_n \omega_n^2 \Omega_n }  \left[\delta (\omega - \Omega_n - \lambda_n ) \pm \delta (\omega - \Omega_n + \lambda_n ) \right] 
\label{Eq3.7}
\end{equation}
where $\Omega_n = \omega_n \sqrt{1+ (\lambda_n/ \omega_n)^2}$ with $\lambda_n = m_n \omega_n^2 \theta / (2 \hbar)$. Finally, the "`random"' force $f_i (t)$ satisfies the conditions:
\begin{equation}
<f_i (t)> =0, \hspace{1 cm} < \left\{f_i (t), f_j (t') \right\}>= \hbar \nu_{ij} (t -t'), \label{Eq3.8}
\end{equation}
where $\left\{A,B \right\} = \left( AB +BA \right)/2$ is the anticommutator and 
\begin{equation}
\nu_{ij} (t) = \int_0^{+ \infty} d \omega \hspace{0.2 cm} \coth \left( \frac{\hbar \beta \omega}{2} \right) \left[ \delta_{ij} I^{(+)} (\omega)  \cos (\omega t) + \epsilon_{ij} I^{(-)} (\omega)  \sin (\omega t) \right] \label{Eq3.9}
\end{equation}
is the noise kernel. As usual, the two kernels are related by the fluctuation-dissipation theorem. Given the kernels $\eta$ and $\nu$ one is able to write an expression for the matrix coefficients ${\bf A},{\bf B},{\bf C},{\bf D}$ of the master equation (\ref{Eq3.4}). So far the expression for these coefficients were derived in the weak coupling limit and for the Caldeira-Leggett model in the Ohmic regime \cite{Dias3}. These results will be presented in a forthcoming publication.

\subsection*{Acknowledgments}

\vspace{0.3cm}

\noindent The work of CB is supported by Funda\c{c}\~{a}o para a Ci\^{e}ncia e a Tecnologia (FCT)
under the fellowship SFRH/BD/24058/2005. The work of NCD and JNP was partially supported by the
grants POCTI/MAT/45306/2002 and POCTI/0208/2003 of the FCT.

\end{document}